\documentclass{article}
\usepackage{amssymb}
\usepackage{makeidx}
\usepackage{amsfonts}
\usepackage{amsmath}
\usepackage{color}
\usepackage{graphicx}
\usepackage{hyperref}

\newtheorem{corollary}{Corollary}

\newtheorem{lemma}{Lemma}

\newtheorem{proposition}{Proposition}

\newtheorem{assumption}{Assumption}

\newenvironment{proof}[1][Proof]{\noindent\textbf{#1.} }{\ \rule{0.5em}{0.5em}}

\begin{document}

\title{Deliberation Among Informed Citizens\\ \vspace{0.2cm} \large{- The Value of Exploring Alternative Thinking Frames -}
\vspace{0.3cm}
}

\author{Ariane Lambert-Mogiliansky\thanks{%
Paris School of Economics, 48 Boulevard Jourdan, Paris 75014.
(alambert@pse.ens.fr)} \thanks{%
I thank Fran\c{c}ois Dubois et Zeno Toffano for very useful discussions and suggestions in the early stages of the project. } and
 Ir\'{e}n\'{e}e Fr\'{e}rot \thanks{%
Laboratoire Kastler, Brossel, Sorbonne Universit\'{e}, CNRS, ENS-PSL
Research University, Coll\`{e}ge de France, 4 Place Jussieu, 75005 Paris,
France. (irenee.frerot@lkb.upmc.fr)}}
\maketitle

\begin{abstract}
In this paper we investigate the potential of deliberation to create consensus among fully informed citizens. Our approach relies on two cognitive assumptions: i. citizens need a thinking frame (or perspective) to consider an issue; and ii. citizens cannot consider all relevant perspectives simultaneously, they are incompatible in the mind. These assumptions imply that opinions are intrinsically \textit{contextual}.

Formally, we capture contextuality in a simple quantum-like cognitive model. We consider a binary voting problem, in which two citizens with incompatible thinking frames and initially opposite voting intentions deliberate under the guidance of a benevolent facilitator. We find that when citizens consider alternative perspectives, their opinion may change. When the
citizens' perspectives are two-dimensional and maximally uncorrelated, the probability
for consensus after two rounds of deliberation reaches 75\%; and this probability increases proportionally with the dimensionality (namely, the richness) of the perspectives. When dealing with a population of citizens, we also elaborate a novel rationale for working in subgroups.

The contextuality approach delivers a number of insights. First, the diversity of perspectives is beneficial
and even necessary for deliberations to overcome initial disagreement. Second, successful deliberation demand the active participation of citizens in terms of
``putting themselves in the other's shoes". Third, well-designed procedures managed by a facilitator are necessary to secure increased probability for consensus. A last insight is that the richness
of citizens' thinking frames is beneficial, while the
optimal strategy entails focusing deliberation on a properly reduced problem.

\end{abstract}

\section{Introduction}

Recently representative democracy has been questioned and is widely
perceived as being in crisis in most developed countries. At the same time,
more participative forms of democracy are gaining interest (\cite{Science19}%
). They are most often framed as direct democracy (in restricted contexts)
and based on sortition (in larger contexts) and they always rely on
deliberations. "Deliberative democracy is now arguably the main theme in
both democratic theory and the practice of democratic innovations"(\cite%
{niemeyer24}). Theories of collective decision-making are traditionally
partitioned into two major fields: those dealing with issues related to the
aggregation of diverse preferences (Social choice) and those dealing with
citizen's active participation aiming at fostering reciprocal understanding
and compromise and move toward consensus(see \cite{list2018} for a review)%
\footnote{%
Deliberative institutional experimentation is flourishing throughout the
world (a catalog is available at https://participedia.net/ ).}.

Social choice theory is repleted of impossibility theorems which have their
foundations in the early work of Condorcet about voting cycles which was
generalized by Arrow (\cite{sen14}) fundamental results showing that
every voting mechanism will fail to satisfy at least one among widely
accepted conditions of fairness and rationality.\footnote{%
When there are at least three alternatives and the domain of preferences is
unrestricted, the following axioms are inconsistent: Pareto Efficiency,
Independence of Irrelevant Alternatives, Non dictatorship, and Social
Rationality.} According to Jon Elster (\cite{Elster05}), social choice
theory view the political process as instrumental and the decisive political
act is a \textit{private} action, an individual secret vote. The goal of
politics is the optimal compromise between \textit{given} and irreducably
opposed private interests. Social choice theories investigate the properties
of aggregation rules and procedures. In contrast other views deny the
private character of political behavior and its pure instrumental nature.
According to Jurgen Habermas the goal of politics should be rational
agreement and the decisive political act is that of engaging in public
debate with the objective of reaching consensus. For participatory democrats
from John Stuart\ Mills to Carole Pateman, the goal of politics is the
transformation and education of participants so that politics is an end in
itself (\cite{Elster05}). Data from Deliberative Polls support the
hypothesis that people do change their opinion and this happens not only
under the impact of better information (\cite{Listetal13}; \cite{Farrar10}).
J. Dryzek writes that a \textquotedblleft defining feature of deliberative
democracy is that individuals participating in democratic processes are
amenable to changing their minds and their preferences as a result of the
reflection induced by deliberation.\textquotedblright\ (\cite{dryzek02})

Deliberative democratic theory is founded on the basic principle that
\textquotedblleft outcomes are democratically legitimate if and only if they
could be the object of a free and reasoned agreement among
equals\textquotedblright\ (\cite{Cohen97}). The question is how does this
process of presenting arguments leads to agreement among equals. Some
scholars have argued that reasoned public deliberation lends legitimacy
because the proposals that are sustained and survive through the process of
deliberation are simply better in terms of their overall quality. This way
of explaining the value of public deliberation and its connection to
political justification presumes that there exist some procedure independent
criteria of rightness or correctness. Many epistemic democrats (\cite%
{estlund97}; \cite{lande13}) hold this view. Other have proposed that
post-deliberation outcomes are more justified than simple non-deliberative
aggregated outcomes because the very procedure of reasoned public
deliberation embodies or manifests core values of basic human morality and
political justice, and it forces participants to be attentive toward the
common good (\cite{Christ97}; \cite{Cohen97}; \cite{Knight97}; \cite{Rawls97}%
). Finally, a number of scholars have argued that reasoned public
deliberation may also complement (or even nullify the need for) aggregative
voting mechanisms: by generating unanimous agreement; by \textquotedblleft
inducing a shared understanding regarding the dimensions of
conflict\textquotedblright\ (\cite{Knightjohn94}); or by inducing
\textquotedblleft single peaked preferences\textquotedblright\ among the
voters, which prevents majority rule from generating majority cycles (\cite%
{dryzeklist03}; \cite{Listetal13}). Bohman emphasizes both the
transformative and epistemic benefits of confronting a diversity
perspectives in deliberations.\cite{Bohman06} Our approach is, in its
spirit, close to that of Bohman's. We view the process of deliberation as a
procedure that invites citizens to explore alternative perspectives.

The central hypotheses of this paper is that i. to be able to consider an
issue, people have to build a representation of that issue. Building a
representation requires selecting a perspective or a thinking frame (a
model), ii. there exist perspectives that people cannot consider
simultaneously, they are incompatible in the mind. This has the crucial
implication that no single perspective can aggregate all relevant
information: opinions are contextual. In a close spirit Niemeyer et al.
write "Deliberative reasoning as we characterize it, recognizes the
possibility of identifying the set of relevant considerations, while falling
short by failing actively to take all of them into account to capture the
complete picture" (\cite{niemeyer24} p. 347). To focus on the evolution of
opinions due to their contextuality, we consider deliberation exclusively as
a process of confronting alternative models with no improvement in
information. Kinder (2003) writes: "frames supply no new information.
Rather, by offering a particular perspective, frames \textit{organize - or
better reorganize - }information that citizens already have in mind"\cite%
{kinder03}. Frames suggest how politics should be thought about, encouraging
citizens to think in particular ways"\footnote{%
This reminds of Aragones et al. \cite{aragones95}. They establish that
finding the best functional rule to explain a set of data is an NP-complete
problem. Fully informed people may have different, mutually incompatible,
predictions without it being possible to establish that some prediction is
better than another. The truth of a prediction depends on the correlation
that is being investigated i.e., it is "contextual".} To address the
contextuality of opinions, we turn to the most widely recognized formal
approach that features the co-existence of alternative representations of
one and the same object: the Hilbert space model of Quantum Mechanics (QM)%
\footnote{%
In quantum Physics, the property values that define a system depend on how
you measure them, the system is contextual to the measurement
apparatus.There exists however of stable objective truth in the larger space
that includes the system and all possible measurements which is formalized
in the Hilbert space model of QM.}. Under the last decades, quantum-like
models of contextuality have been developed in Social Sciences to explain a
variety of behavioral anomalies (for overviews, see refs.~\cite%
{bubu12}, \cite{Yearbu16}, \cite{Jerry20}) We briefly
introduce the quantum cognition approach in section 1.2. We immediately
reassure the reader that no prior knowledge of QM or Hilbert space is needed
to read the present paper.

To fix idea consider a collective decision issue related to the introduction
of Individual Carbon Budget (ICB). One perspective or thinking frame is the
environmental one i.e., ICB properties to reduce green house gas (GHG) emissions
. Another thinking frame relates to individual liberties i.e., ICB's
impact of individuals' freedom of choice. People can very well consider
those two perspectives in sequence. However, they may have difficulties to
consider them simultaneously. As a consequence, they cannot aggregate
information from the environmental and the liberty frames \textit{in a
stable way }instead a citizen's opinion depends on the order in which the
two (incompatible) perspectives are explored (probed). This instability is
the expression of\textit{\ intrinsic} contextuality i.e., an opinion does
not exist independently of a frame (its context) and that it is impossible
to integrate all frames into a unique "super frame"\footnote{%
Instead, there exists an irreducible multiplicity of alternative frames that
are equally valid to characterize an issue. In physics this is a defining
feature of sub-atomic particle i.e., some properties are Bohr complementary
and cannot have a definite value simultaneously. In psychology and social
sciences this is proposed as a defining feature of mental constructs like
the representation of an issue.}.

There exists, to the best of our knowledge, no formal approach that features
the impact of the variety of thinking frames on the outcome of
deliberations. However, recent theoretical and experimental works on quantum
persuasion by Danilov et al.(\cite{dan18a}, \cite{dan18b}) can be relevant
to the context of deliberation. In line with Danilov et al., we view a
person's opinion as a non-classical (quantum) object characterized by its
state. Alternative thinking frames are modelled as alternative basis
of the opinion space. The deliberative process is modelled
as a structured sequence of measurements i.e., an ordered sequence of
"exploration" of different perspectives by the citizens. By force of the
intrinsic contextuality of opinion, measurements move the opinion state in
non-deterministic manner that reflects the correlations between the bases (the
thinking frames). Given this technology, we investigate the properties of
procedures that satisfy some requirements of deliberation put forward in the
political philosophical literature. Deliberation should be a respectful,
open-minded, fair and equitable communicative process which aims at
achieving maximal consensus with respect to the decision at stake.\ 

We consider a setting where there is a Yes or No$\ $decision to be made%
\footnote{%
The voting issue is two dimensional implying that we are not addressing
the impossibility theorems of collective choice (Arrow 1951).}.
The objective of deliberation is to achieve democratic legitimacy in the
following sens. First, the procedure should give a fair chance to
everyone's opinion to affect the decision. Second, full consensus is the
overarching goal which translates in maximizing the support for the final
decision. In this context, the paper addresses two central questions: 1. How
can (fact-free i.e., without additional information) deliberation affect
citizens' voting behavior? 2. How should we structure deliberation to maximize
the probability for consensus? A central assumption is that citizens are willing to explore
alternative thinking frames before deciding how to vote. These alternatives
frames can be provided by the citizens themselves and/or experts. The
procedure is managed by a facilitator. We find that in the two-person case when
starting from opposite voting intentions, fact-free deliberation procedure
always achieves some extent of consensus. The largest chance of consensus
obtains when the citizens' perspective are maximally uncorrelated\footnote{Perspective A and B
are said to be distant (uncorrelated) whenever a citizen's opinion in
perspective A has no(little) predictive power for her opinion in perspective
B.}. For the case the perspectives are two-dimensional
consensus can be reached with probability $3/4$.
The results generalize to n-dimensional perspectives where we show that
the more fine-grained the opinion, the easier to reach consensus. A first central insight is that citizens' willingness to seriously consider an alternative perspective by "putting themselves in
someone else's shoes" can have a very significant impact on opinions. And
secondly the more "distant" the considerations of the other citizen from the
own perspective, the more powerful the impact. We next extend to a
population of citizens and to multiple perspectives. An important finding is
that to maximize the probability of convergence to consensus, the
facilitator tends to discriminate between citizens i.e., in each round some are active while others remain passive. In addition, we find that it may be optimal to target the standing minority option as the projected consensus. In the Discussion section
we address epistemic considerations and the performance of our model with
respect to the ideal pf deliberative democracy.

The main takeaways from our account of
thinking frames in the analysis of fact-free deliberations are: i. The diversity of
thinking frames among citizens not only is no obstacle but a necessary
condition for deliberation to deliver consensus; ii. Deliberations can exhibit a powerful
transformative power that hinges on a true willingness of participants to explore other thinking frames; iii. Well-designed procedures
monitored by a facilitator are needed to move toward consensus.

This paper contributes to the literature on the value of pre-voting
deliberation by providing a formalisation of opinion formation appealing to
the intrinsic contextuality of opinions. Most formal approaches to deliberation belong the epistemic tradition
which postulate a single (common) correct decision. Among the most recent
ones Dietrich and Spiekermann introduce some behavioral dimensions (sharing
and absorbing) to the information theoretic approaches (see e.g., \cite%
{Dietrich24}). The other formal strand of literature is game-theoretic which
emphasize incentive to share of withhold information (see e.g., \cite{piva21}%
). The (quantum) contextuality revolution recasted the issue of objective
truth and knowledge as witnessed by the wealth of the epistemological
literature under the last century (see e.g., \cite{despa12}).\ Our approach
based on the most standard formalisation of intrinsic contextuality is
closely related to Bohman's experiential perspective approach in \cite%
{Bohman06}. He emphasizes the transformative and epistemic benefits of
confronting the diversity perspectives in deliberations. We provide a formal
mechanism for this transformation and derive procedures managed by a
facilitator to increase consensus. Because our deliberation involves no
improvement in information, it is complementary to classical epistemic
approaches. Because intrinsic contextuality precludes the uniqueness of truth, it bring us close to procedural approaches to deliberation
(see e.g., \cite{beauvais16}) which recognize a value to deliberations
in terms of actualizing basic values of
democracy. Our results also contribute with new results with respect to the 
tension between efficiency and legitimacy within the process of deliberations itself \cite{dryzek01}.

\section{Contextuality}
In this section, we illustrate with a story the kind of situation we have in mind and introduce the quantum cognitive approach.   
\subsection{A story}

A community needs to decide whether or not to
introduce an Individual Carbon Budget (ICB) scheme. To
be able to relate to the issue, citizens build a mental representation using a thinking frame. Various aspects are of relevance to citizens, environmental
efficacy, impact on individual liberties, legal feasibility etc... each corresponding to specific thinking frames. The citizens participating in the deliberations are open-minded. They are willing explore frames alternative to their own. An
environmentalist discovers how she thinks about ICB in terms of individual
liberties etc... As a result of exploration, the citizens discover that their
initial opinion has evolved. They have not learned new facts but experiencing a new
frame i.e., "putting themselves in the shoes of someone who thinks in
different terms" modifies their state of mind. After some rounds of (guided)
deliberation, they find that they agree to a significantly larger extent.

\subsection{Quantum cognition}

\begin{figure}
    \centering
    \includegraphics[width=0.5\linewidth]{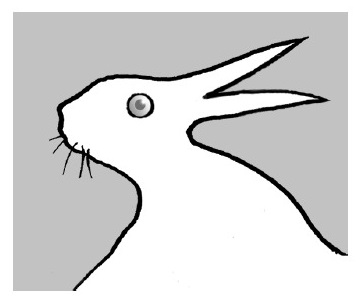}
    \caption{What do you see?}
    \label{fig_lapin}
\end{figure}

It is a common place that human beings are not capable of holding very
complex pictures in mind. We consider reality focusing on one perspective (or thinking frame) at
a time and show difficulties in combining perspectives in a stable way. This
inability to seize reality in its full richness suggests that the process of
developing an understanding of the world may not look like a puzzle that is
assembled progressively. Instead, the human mind may exhibit structural
"limitations" in terms of the incompatibility of perspectives in a way
similar to properties in quantum mechanics. Ambiguous pictures of the kind
provided in Fig.~\ref{fig_lapin} provides a suggestive illustration of this phenomenon.
You may see a duck or a rabbit (you may oscillate between the two as for the
Necker cube) both are correct but you cannot see both simultaneously.

To many people it may appear artificial to turn to quantum mechanics (QM)
when investigating human behavioral phenomena. However, the founders of QM,
including Bohr and Heisenberg recognized an essential similarity between the
two fields\footnote{%
In particular, Bohr was influenced by the psychology and philosophy of
knowledge of Harald H\"{o}ffding (see Bohr (1991) and the Introduction in
Bitbol (2009) for an insightful discussion).}. In both fields the object of
investigation cannot (always) be separated from the process of
investigation. QM and in particular its mathematical formalism was developed
to respond to a general epistemological challenge: how can one study an
object that is being modified by the measurement of its properties? QM is a
general paradigm for intrinsic contextuality (i.e., non-separability between
the object and the operation of investigation). It should, therefore, be
viewed as truly legitimate to explore the value of the mathematical
formalism of QM in the study of human behavioral phenomena - without
reference to Physics.

The quantum paradigm has been proposed in decision theory and psychology to
describe preferences, beliefs, attitudes, judgements and opinions (see \cite%
{bubu12}, \cite{Yearbu16}, \cite{dan18c},\ \cite{Jerry20} among others). The
approach has allowed providing a unified framework that accommodates a large
variety of, so called, behavioral anomalies: preference reversal,
disjunction effect, cognitive dissonance, framing effects, order effects
etc. (see e.g. \cite{ALMZZ09} and for empirical work \cite%
{emp09},\cite{calm21}).

In this paper, we recognize that our representation of the world is a mental
object (a representation) which may exhibit non-classical features and we
derive some implications of this hypothesis for the dynamics of opinion
formation in the process of deliberation. The most important element that we
borrow from QM is the notion of Bohr-complementarity applied to mental
perspectives\footnote{%
It parallels the Bohr complementarity of properties for sub-atomic particles
e.g., spin along different angles.}. In line with QC, we propose that Bohr
complementarity of perspectives captures the cognitive limitations that are
responsible for our difficulties to syntheses information along different
perspectives into a single stable picture. Just as in QM, the system (here our
mental picture) makes discrete jumps when attempting to find a determinate
value along distinct incompatible perspectives.

Recently theoretical and experimental applications to persuasion have shown
how fruitful this approach could be (\cite{dan18a}, \cite{dan18b},\cite%
{calm21}). This paper is in continuation with those works. A citizen's
opinion is a non-classical (quantum) object characterized by its state.
Alternative thinking frames are modelled as alternative basis (incompatible
properties) of the representation of the decision issue. Deliberation amounts to a sequence of
measurements (probing arguments). Measurements moves the opinion state in a
non-deterministic way that reflects the correlations between the bases (the
thinking frames). With this formal description of the process of
deliberation, we investigate procedures that satisfy some requirements on
deliberation put forward in the literature. 

We emphasize that the quantum cognition approach does not assume a quantum
physical nature to the determinants of our opinions. Neither do we dwell
into the psychology or neurology of the transformation of belief/opinion and
preferences\footnote{%
The approach is an abstract way of capturing the fact that experiencing
another perspective may have neurological, emotional and other impacts with
consequences for opinion.}. A presumption is that the correlations between
perspectives which structure the mind exhibit some extent of regularity
across individuals. As in Physics, the quantification of correlations is an
empirical question.

\section{Model}

\subsection{Basic structure}

We shall formulate our model of deliberation in terms of a mediated
communication game. There is a set $\Omega $ of deliberator-citizens who may
alternatively be in one of two roles R (Receiver) and S (Sender), one
facilitator and a pool of experts. The interaction between the citizens and
the facilitator is simplified by assumption 0 below. We are interested in
deliberation aimed at influencing Receivers' vote over uncertain options
which we model as quantum lotteries following Danilov et al. 2018 (\cite%
{dan18c}). The formal model shares significant features with the quantum persuasion
model (\cite{dan18a}, \cite{dan18b}).\ 

\begin{assumption}
\label{assum_0}
Citizens taking part in deliberation are willing to explore alternative thinking frames. They follow the recommendations of the facilitator.
\end{assumption}

Citizens participating in deliberation are willing to engaged in real mental experiences with an open mind. This capture their respect for each other and their trust in the procedures as a legitimate way to reach
a collective decision. The facilitator's recommendations relates exclusively
to invitations to present an argument and invitation to probe a perspective
(see below for definition). In this paper, we are not addressing possible
incentive issues related to these operations. Instead, we consistently with
the normative literature on deliberation assume goodwill from the side of
citizens. \medskip

The description of a quantum system starts with the fixation of a Hilbert
space $H$ (over the field $\mathbb{R}$ of real numbers or the field $\ $of
complex number). Physicists usually work with the complex field $\mathbb{C}$%
. We, for simplicity, shall work with the real field $\mathbb{R}$, although
everything goes with minor changes for the complex case. Let
(\textperiodcentered , \textperiodcentered )\ denote the scalar product in
Hilbert space $H\ $(in our case a finite dimensional space).

We shall be interested not so much in the Hilbert space as in operators,
that is linear mappings $A:H\rightarrow H$. Such an operator $A$ is
Hermitian(or symmetric over $\mathbb{R}$) if $(Ax,y)=(x,Ay)\ $for all $%
x,y\in H$. A Hermitian operator $A$ is non-negative if $(Ax,x)\geq 0\ $for
any $x\in H$\footnote{%
General Hermitian operators play the role of classical random variables. In
fact for any Hermitian operator $A_{{}}\ $and opinion $O,\ $we can define
the `expected value' of $A$\ in state $O\ $as ${\rm Tr}(AO)$.The expected utility
of voting action $yes$ (to ICB) are presented by operator $Y$ in an opinion
state $O$ is expressed as ${\rm Tr}(YO$)\ and this number linearly depends on $O$.
In this way, Receiver's preferences over voting options are determined by
her opinion state $O$.}.

\subsection{Opinion State and perspectives}

Each individual $j$ is characterized by her opinion state $O^{j}$ and her
thinking frame, operator $P^{j}$ (perspective).\medskip

\textit{Opinion state}

With the help of the trace one can introduce the notion of state of a
quantum system. The trace ${\rm Tr}$ of a matrix can be defined as the sum of its
diagonal elements. It is known that the trace does not depend on the choice
of basis. Let $o$ (for opinion) be an element of $H$ with length 1 (that is $%
(o,o)=1$). Let $P_{o}$ be the orthogonal projector\footnote{%
A projector is an Hermitian operator such that $P^{2}=P)$} on $o$, that is $%
P_{o}(x)=(x,o)o$ for any $x\in H,\ {\rm Tr}(P_{o})=(o,o)=1$, therefore $P_{o}$ is
a state also denoted by operator $O(=P_{o})$. Such states are called pure,
we shall be exclusively dealing with pure states\footnote{%
This is by distinction with mixed states. Pure state are complete
information state but because of intrinsic indeterminacy, the values along
alternative (incompatible) perspectives can never be sorted out
simultaneously, they remain stochastic.}. The non-negativity of the operator 
$O$\ is analogous to the non-negativity of a probability measure, and the
trace 1 to the sum of probabilities which equals 1. This means that an
opinion state is formally identical to a (subjective) belief state \cite{dan18b}.\\

\textit{Perspectives}

The formal account of thinking frames is a key building block of our theory. It is intimately linked with our cognitive assumptions: 
\begin{assumption} ~
     \begin{enumerate}
    \item[i] Citizens cannot address reality immediately. They need to build a \textit{representation} of the voting issue using a thinking frame;
    \item[ii] Citizens cannot resort to a "super frame" that aggregates all relevant aspects.
\end{enumerate}
\label{assum_cog}
\end{assumption}

Assumption 2 implies that, for the citizens, the voting issue admits a finite number $k$ of 
equally valid (Bohr complementary) thinking frames, the \textit{perspectives}. Formally, there is a set $\mathcal{P}$\ of perspectives (or thinking frames, we use the
terms interchangeably). A perspective $P=\left( P_{1},...,P_{n}\right) ,\
P_v:H\rightarrow H\ \ $is a Hermitian operator. We focus on a limited subset
of operators referred to as direct (or von Neumann) measurements. These
simple devices are sufficient for the purpose of the present paper. Such a
device is given by a family of projectors ($P_{v},v\in V,\ $ $V$ is the set
of values of the device.) with the property $\sum P_{v}=E$, where $E\ $is
the identity operator on $H$. The probability $p_{v}$ to obtain outcome $v\ $%
(in a state $o\in St$) is equal to ${\rm Tr}(P_{v}O)$ and if we obtained $v,\ $the
revised opinion-state is $O_{v}=P_{v}\ $ because projector $P_{v}$ is
one-dimensional (that is, a pure state). If we repeat the measurement, we
obtain the same outcome $v$ and the state does not change\footnote{%
This type of measurement is repeatable or \textquotedblleft first
kind\textquotedblright .}. An important feature that we want to emphasize is
that the expected revised opinion $O_{ex}=\sum_{v\in V}P_{v}O_{v}=\sum_{v\in
V}P_{v}OP_{v}\ $is generally different from the initial opinion-state $O$.
That is, although the opinion state has the structure of a probability
distribution, the revised opinion in the non-classical context is \textit{not%
} subject to Bayesian plausibility as noted in \cite{dan18b}. This feature
plays an important role in the analysis.\medskip

\textit{Probing a perspective }

Probing a perspective, is formalized as an operation whereby one applies
some perspective $P$ to an opinion state $O.$ In practical terms, it
corresponds to questioning oneself in the terms of perspective $P$ i.e.,
which specific value do I agree with? We focus exclusively on \textit{%
complete measurements} that is probing operations that fully resolves
uncertainty with respect to the $P$ perspective. The outcome of probing a
perspective is one of its eigenvalues.

\begin{figure}
    \centering
    \includegraphics[width=1\linewidth]{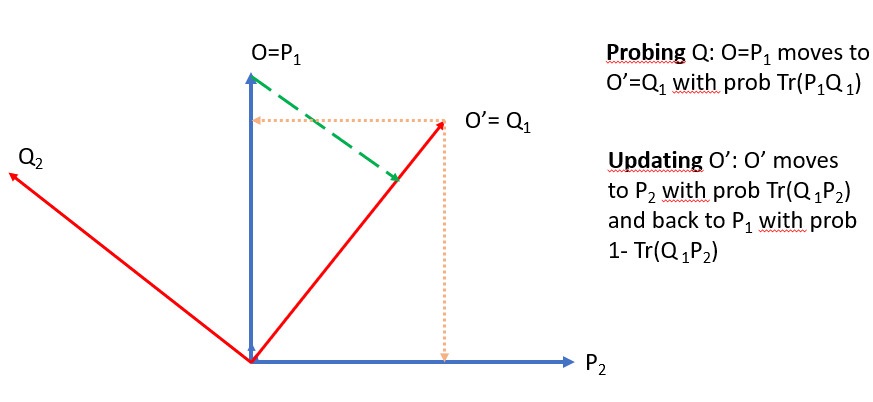}
    \caption{Probing perspectives}
    \label{fig_schema}
\end{figure}

Assume citizen $j$ is characterized by $O\ $and$\ P^{j}$ such that $%
O=P_{i}^{j},\ $probing an argument in an \textit{alternative} (i.e.,
incompatible) perspective $Q$ with arguments $Q_{i}$ is called \textit{%
challenging one's opinion}. Challenging an opinion with $Q\ $generates an
new state $O^{^{\prime }}\in \left\{ Q_{1},...Q_{n}\right\} $ and, as a
consequence, it upsets her previously held value $v_{i}\ $in $P^{j}$%
\footnote{%
In quantum mechanics, it is postulated that the state (of a measured system)
changes in accordance with the von Neumann--L\"{u}ders postulate. More
precisely, a system that was in state $O$ transits to the state $O^{\prime
}=POP/{\rm Tr}(POP$) as a result of performing a measurement that yields event $P$%
.\ In our simplified setting, we say that the opinion state transits into
(pure state) $P.\ $}. For instance the individual is confronted with the
liberty perspective and reflects over whether she thinks that ICB is
contrary to fundamental individual liberties $Q_{1}$ or implies acceptable
limitations $Q_{2}$ or has no implication for individual liberty $Q_{3}$.
After the process of probing, the revised opinion is either $Q_{1},Q_{2}$ or 
$Q_{3}.$ A central feature of the model is that if the citizen started from $%
O_{{}}^{{}}=P_{2}^{j}$ so she thinks/believes ICB are environmentally useless%
$,$ probing argument $Q_{{}}$ and obtaining for instance $i_{Q}=1$ changes
her previously held opinion into $Q_{1}.\ $So if the citizen probes $P\ $%
anew ${\rm Tr}\left( Q_{1}P_{i}\right) >0$ for $P_{i}\neq 2$, the citizen has
changed her mind from initial $P_{2}\ $and now believes (with positive
probability) that ICB are environmentally valuable $P_{1}$.\ This is the
crucial property that
generates a potential for opinions to evolve without additional information.
This is illustrated in Fig.~\ref{fig_schema} in the two-dimensional case. With initial
opinion state given by $P_{1}$ ,our citizen probes the liberty perspective $%
Q $ and with probability ${\rm Tr}\left( P_{1}Q_{1}\right) $ she finds that in
that perspective her opinion is $Q_{1}$ (and with the complementary
probability $Q_{2}).$\ She then update her opinion (to be able to evaluate
the voting options see below) and with probability ${\rm Tr}\left(
Q_{1}P_{2}\right) $ she does not recover her initial opinion instead she now
holds opinion $P_{2}.$ 

Throughout, we use the term \textit{probing} for a measurement of an \textit{%
alternative} perspective to the citizen's own perspective. While \textit{%
updating} refers to the measurement operation of the \textit{own}
perspective.

\subsection{Utility and voting\textit{\ }}

Voting is binary, a Yes or No choice. Citizens are endowed with preferences
that allow evaluating the utility value of the two options given their
individual opinion state. .

\begin{assumption}
Citizen $j\in \Omega $ exclusively attributes utility to voting options in
one of the eigenstates of her own frame i.e. in some state $O^{j}\in \left\{
P_{i}^{j}\right\} ,$\ where $P_{{}}^{j}$\ is her \textit{own} perspective. 
\label{assum_util}
\end{assumption}

Assumption \ref{assum_util} captures a central feature of thinking in frames. A citizen is
only capable to evaluate her utility in terms of her own perspective. Citizens can explore other perspectives and adopt any
possible opinion. However, an opinion state can guide voting \textit{only}
when formulated in terms of the own perspective. The frame is an essential
part of the citizen's identity, it is the language in which she can
formulate the value of the voting options and make decisions.

Formally, consider a citizen with perspective $\mathcal{P}=(P_{1},...,P_{n})$%
. Her utility "function" is represented by a tuple $\left\{
u_{1}^{Y},..,u_{n}^{Y},u_{1}^{N},...,u_{n}^{N}\right\} $ which associates a real number to option Yes respectively No to each possible opinion state. Generally, for
any arbitrary opinion state $O$, we can formulate the expected utility for the two
voting options as a "quantum lottery" (see \cite{dan18c}), with the
probability for each of the states $P_{i}$ given by ${\rm Tr}(OP_{i})$. The
general formula for the expected utility is then 
\begin{equation}
\mathrm{EU}\left( Y;O\right) =\sum_{i}\mathrm{Tr}\left( OP_{i}\right)
u_{i}^{Y}  \label{EUdef}
\end{equation}%
and similarly $\mathrm{EU}\left( N;O\right) =\sum_{i}\mathrm{Tr}\left(
OP_{i}\right) u_{i}^{N}$. When $\mathrm{EU}\left( Y;O\right) \geq \mathrm{EU}%
(N;O)$, the citizen prefers the Yes option. Note that in order for the
expression in Eq.
(\ref{EUdef}) to guide voting behavior the citizens must be able to
compute $tr\left( O_{{}}^{{}}P_{i}^{{}}\right) $ for any $O.\ $This is very
demanding. In particular, it requires that citizens have knowledge about the
correlations between all alternative perspectives and their own. We do not
make that assumption, instead whenever in $O\neq P_{i}^{{}},$ she will have
to "go back" to her own thinking frame before voting (see below).

Citizens' utility for the voting options defines two complementary subsets: 
\begin{eqnarray*}
\mathcal{Y} &\equiv &\left\{ P;\ \mathrm{EU}(Y;P)\geq \mathrm{EU}%
(N;P)\right\} \\
\mathcal{N} &\equiv &\left\{ P;\ \mathrm{EU}(Y;P)\leq \mathrm{EU}%
(N;P)\right\} .
\end{eqnarray*}%

\begin{assumption} 
 A citizen is endowed with a thinking frame that allows separating between voting options, namely $\mathcal{Y}$ and $\mathcal{N}$ are nonempty.
 \label{assum_diff}
\end{assumption}

Assumption \ref{assum_diff} excludes citizens whose voting decision is fixed and
therefore cannot be affected by deliberation. For any citizen $j,$ the own
thinking frame $P^{j}$, allows separating among voting alternatives. 

The voting behavior is most simple. We know by Assumption \ref{assum_util} that citizen $j\ $must
be one of her perspective's eigenstates  say $P^{j}$
then if $u\left( Y;P_{i}^{j}\right) \geq $ $u\left( N;P_{i}^{j}\right) ,$
citizen $j$ cast a Yes vote otherwise she casts a No vote, i.e., voting is
sincere (non-strategic). The winning option is the one that has obtained the
largest number of votes.

\subsection{Deliberation}

Deliberation is modelled in terms of a multiple round mediated communication
game. The facilitator is the sole true player in this game. By Assumption \ref{assum_0}
the citizens follow the facilitator's recommendations. In each round, he
chooses an argument $P_{i}^{k}\ $that will be exposed by citizen $k$ and a selection of citizens (defined as
being \textit{active}) whom he invites to probe the corresponding perspective 
$P^{k}$. All other citizens remain "passive", they only listen. When hearing an
argument, an active citizen not only decides for herself whether or not she
agree with the specific argument $P_{i}^{k}\ $, she explores $P^{k}$ by
thinking in its terms so as to determine which argument
she agrees with. Formally, the probing operation is a complete measurement resulting in a
complete information (pure) state i.e. one of the eigenstates of  $P^{k}$%
\footnote{%
This is more demanding than simply deciding to agree or reject the presented
argument. Indeed if the argument is rejected, the revised state will not be
an eigenstate of the perspective under exploration but a superposition of states.}.
After challenging her opinion by probing $P^{k}$, the citizen "updates" her
opinion by reassessing her position in her own perspective where she can
evaluate the utility value of voting options.

\textit{The facilitator strategy}

In each round the facilitator chooses two things: a perspective $P\in 
\mathcal{P}\ $($\mathcal{P}$ is the set of all perspectives including the
experts') and $\omega \subset $ $\Omega $, where $\Omega $ is the set of all citizens, who are
called to be active. The facilitator's objective is to maximize the support
for the most supported voting option. Let $\Omega ^{Yt}$ and $\Omega ^{Nt}$
be the subsets of citizens that support the Yes respectively the No vote in
period $t,\ $with $\left\vert \Omega ^{Yt}\right\vert (\left\vert \Omega
^{Nt}\right\vert )=y^{t}\left( n^{t}\right) $\ the number of Yes(No)
supporters. Deliberation is a finite period game aimed at maximizing the
probability for consensus among citizens before period $T,T\geq 1.$
While consensus in larger groups is difficult to reach, it can be
approached. We operationalize this with a "score function" $s^{t}\left(
.\right) =\max \left\{ y^{t},n^{t}\right\} ,\ s^{t}\left( .\right) $\ is
simply the score of the most supported option in period $t$. Next, in a
multi-period game a strategy must generally describe what to do in each
period $t\leq T-1\ $after each possible history. A first crucial remark is
that the vector of opinions $\mathbf{O}^{t}\mathbf{=(}%
P_{{}}^{1t},...,P_{{}}^{Nt})$\ captures all relevant information about
history at time $t$. This follows from the fact that the probability for a
citizen's opinion change, ${\rm Tr}(OP_{i}),\ $only depends on the current state
of opinion - not on history.\ Next, we shall restrict attention to
strategies that maximize the expected score \textit{in each period}. This is
not fully without loss of generality as we discuss in Section  \ref{sec_3P}. Hence, a
strategy is defined as a function from a vector of opinions $\mathbf{O}^{t}$
to a pair $\left( P,\omega \right) $ including a perspective $P$ and a set
of active citizens $\omega :$ $\mathbf{O}^{t}\rightarrow \mathcal{P}\times
\left( 2^{N}-1\right) $.\ The facilitator objective function in each period $%
t\leq T-1$ 
\begin{equation}
\max_{P\in \mathcal{P},\omega \subset \Omega }Es^{t}\left( P,\omega
;\mathbf{O}^{t-1}\right) =\max_{P\in \mathcal{P},\omega \subset \Omega } \left\{ \max \left\{ Ey^{t},En^{t}\right\} \right\}
\label{ObjFac}
\end{equation}%
where $\mathbf{O}^{t-1}$ is the vector of opinions inherited from the previous period
and $Ey^{t},En^{t}$ is the expected number of citizen who support the Yes
respectively No vote at the end of the $t$ round. We note that the
formulation in Eq.~\eqref{ObjFac} implies that facilitator has no preference over voting
options\textit{. }The facilitator's indifference is a non-innocuous
assumption. We leave issues related to safeguards against possible
manipulation for further research.

We make the following informational assumption:

\begin{assumption}
The deliberation facilitator has knowledge about all perspective and the correlations between them.
\end{assumption}

The facilitator has an informational advantage compared to citizens in terms
of knowing the full structure of the opinion state space. He is aware of all
possible perspectives and how they correlate to each other. This is the
critical resource that allows to him to optimize the deliberation process.
We also assume that he has access to a pool of experts who can present arguments from any possible perspective $P^{e}.$ Experts are simply "tools" that the
facilitator can call upon whenever he wants. This implies that the facilitator is not constrained by the
perspectives represented in the population of citizens. However, since the spirit
of deliberation is (also) to give voice to citizens, we shall give
particular attention to what can be achieve without appealing to experts.

\section{Analysis}

For the ease of presentation we start the analysis with deliberation between
two citizens. This allows to derive basic results before moving to the case
when deliberation involves larger groups of citizens.

\subsection{2 citizens}

We have 2 citizens Alice and Bob who face a binary collective decision about e.g.,
the introduction of a Individual Carbon Budget scheme. There exist two
relevant aspects: environmental efficacy and individual liberty. These 2
aspects cannot be considered simultaneously by our citizens, they are
assumed incompatible in the mind.

The perspectives are $n-$dimensional, the larger $n,$ the finer the
characterization of the opinions. In term of our lead example, the
environmental perspective can have several possible values (each associated
with its own eigenstate) e.g., ICB is the best solution ot reducing GHG
(greenhouse gas), ICB is one among the better solutions to reduce GHG, ICB
is a good solution etc.. until ICB is worthless to reduce GHG. So expanding
the dimensionality corresponds to allowing for "finer" opinions.

The citizens are endowed with
\begin{description}
\item[i.] An opinion state $O^{A(B)}$;
\item[ii.] An own $n-$dimensional perspective. Denote $A$ the operator
representing Alice's perspective with eigenstates $\left(
A_{1}^{{}},...,A_{n}\right) $ and by $B$ the Hermitian operator representing
Bob's perspective with eigenstates $\left( B_{1},...,,B_{n}\right) $ ;
\item[iii.] A set of utility values associated with the own perspective's
eigenstates: $\left\{
u_{{}}^{A_{1},Y},...,u_{{}}^{A_{n},Y},u_{{}}^{A_{1},N},...,u_{{}}^{A_{n},N}%
\right\} $ for Alice, with $u_{{}}^{A_{i},Y}$ the utility value for Alice
corresponding to the Yes vote in opinion state $A_{i}$, similarly for $%
u_{{}}^{A_{i},N}$ corresponding to the No vote. We define similarly $%
u_{{}}^{B_{j},Y},\ u_{{}}^{B_{j},N},$ $j=1,...,n$ for Bob.\medskip
\end{description}

\textit{Deliberation protocol.--} Before Bob and Alice start deliberating,
they update their own opinion namely they probe their own perspective\footnote{This is most natural since they expect to be invited to expose their opinion}. The
initial opinion states are therefore always eigenstates of the own
perspective: some $A_{i}$ for Alice and some $B_{j}$ for Bob. and similarly
for Bob. We consider a process where the facilitator does not
appeal to experts in the first 2 rounds (we call it the "voice phase").

\emph{Timing}

$t=0$ The facilitator asks for initial voting intentions. And if Alice and
Bob agree, the procedure requires no deliberation, they vote; If they
disagree the first round starts:

\textit{Round 1}

$t=1$\ A random draw determines who will present his/her argument first, say
Alice;

$t=2$\ Alice's argues for Bob in her terms (frame) and Bob\ responds by
probing Alice's perspective;

$t=3\ $Bob updates his opinion by probing his own perspective. If he now
agrees with Alice, they vote. If disagreement remains, the second round
starts:

\textit{Round 2}

$t=4$ Bob exposes his opinion in his own frame and Alice responds by probing
Bob's perspective;

$t=5\ $ Alice updates her opinion by probing her own perspective.

$t=6$ Out of the resulting opinion states if they agree, they vote. If they
do not the process 1-3 is repeated - possibly appealing to experts.

$t=8$\ if disagreement persists at $t=T$, the decision is determined
by a random device (to break the tie).

\subsubsection{The two-dimensional case ($n=2$)}

For the ease of presentation, we start with the case where the citizens'
perspectives are 2-dimensional which allows establishing some results which
we later extend. Since our facilitated communication game proceeds by
(quantum) updating, and the facilitator maximizes the score in each round,
we adopt a forward inductive approach.

Alices's and Bob's respective opinion states are independent so the global
system is characterized by the following 4 possible (combined )
opinion states $\left\{ \left( A_{1}B_{1}\right) ,\left( A_{1},B_{2}\right),\left( A_{2},B_{1}\right) ,\left( A_{2},B_{2}\right) \right\}$\footnote{Formally, we are dealing with tensor products so $A_{1}B_{1}=A\times{B}\in{H^{4}} $}. By
Assumption \ref{assum_diff} and since Alice and Bob only have two possible opinions, we
may, without loss of generality, define $A_{1}$, $B_{1}$ as the opinion
states leading to a Yes vote; and $A_{2}$, $B_{2}$ as the opinion states
leading to a No vote. The voting agreement is then entirely defined by the
fact that either Alice and Bob have respective opinion states $A_{1},B_{1}$,
or they have respective opinion states $A_{2},B_{2}$. In both case the
facilitator's score is maximal and equal to 2. When the vector of opinion states $%
\mathbf{O\notin }\left\{ \left( A_{1}B_{1}\right) ,\left( A_{2},B_{2}\right)
\right\} ,$ to maximize the score, the facilitator needs one of the citizen
to change his or her opinion. Maximizing the expected score is equivalent to
maximizing the probability that this happens. In the two-dimensional case,
the probability for Alice or Bob to change opinion when probing the other's
perspective is entirely governed by a single parameter: $x=\mathrm{Tr}%
(A_{1}B_{1})$. Indeed, as $A_{1}+A_{2}=E=B_{1}+B_{2}$, with $E$ the identity
operator, we have $\mathrm{Tr}(A_{2}B_{1})=\mathrm{Tr}(A_{1}B_{2})=1-x$, and 
$\mathrm{Tr}(A_{2}B_{2})=x$.

We consider deliberation starting from initial disagreement, say Alice and
Bob have respective opinion states $A_{1},B_{2}$. From the facilitator point
of view the 2 two consensus states are fully symmetric Let the initial
random draw give Alice the initiative. The facilitator invites her to
present her argument. Bob is then invited by the facilitator to probe
Alice's perspective, and his opinion state changes to $A_{1}$ with
probability $\mathrm{Tr}(A_{1}B_{2})=(1-x)$, and to $A_{2}$ with probability $%
\mathrm{Tr}(A_{2}B_{2})=x$. Bob then updates his opinion by probing his
own perspective. The probability for reaching consensus is the probability
that Bob's final state is now $B_{1}$, instead of his initial state $B_{2}$.
This state is reached with probability: 
\begin{eqnarray}
\mathrm{prob}(B_{2}\rightarrow B_{1}) &=&\mathrm{Tr}(B_{2}A_{1})\mathrm{Tr}%
(A_{1}B_{1})+\mathrm{Tr}(B_{2}A_{2})\mathrm{Tr}(A_{2}B_{1})  \label{Con2X2}
\\
&=&2x(1-x)
\end{eqnarray}%
Since only Bob's opinion has been challenged, the other consensual state
(namely, $A_{2}$ for Alice and $B_{2}$ for Bob) cannot have emerged.\newline

\textit{Maximally-uncorrelated perspectives.--} Generally, $A$ and $B$
perspectives are said to be maximally uncorrelated when $\mathrm{Tr}%
(B_{i}A_{j})=1/n$ for all $i,j$ (with $n=2$ in the present case, $n$ is the
dimensionality of the perspective), probing the $\mathcal{A}$ perspective
gives Bob equal chance to move into any of the states $A_{i}$ (here, Bob
reaches $A_{1}$ or $A_{2}$ with probability $1/2$). Updating then his
opinion by probing the $\mathcal{B}$ perspective, Bob reaches any of the
opinion states $B_{j}$ with equal probability. Effectively, Bob's initial
opinion state has been completely randomized by the (intermediate) probing of
Alice's completely uncorrelated perspective $\mathcal{A}$. In such a case,
starting from initial disagreement, the chance for reaching consensus after
one round is equal to $1/2$. \ 

The second round following disagreement
proceeds similarly, generating a probability for consensus in states $%
A_{2},B_{2}$ with the same probability $2x(1-x)$. Hence after the two rounds
the probability for reaching consensus is $2x\left( 1-x\right) +\left(
1-2x\left( 1-x\right) \right) 2x\left( 1-x\right) =4x\left( 1-x\right) \left[
1-x\left( 1-x\right) \right] $.

We have the following

\begin{proposition}
\label{Prop 1}
Starting from disagreement on vote between two citizens,

i. fact-free deliberation with fully correlated perspectives has no impact
at all;

ii. with distinct perspectives consensus is reached with strictly positive
probability after a first round;

iii The probability for consensus is largest when the perspectives are
uncorrelated, it reaches $3/4$ after two rounds.
\end{proposition}

\begin{proof}
i. When the two perspectives are fully correlated, we have $\mathrm{Tr}%
(A_{i}B_{j})\in \left\{ 0,1\right\} $: the opinion state are either equal ($%
A_{1}=B_{1}$ and $A_{2}=B_{2}$) or orthogonal ($A_{1}=B_{2}$ and $%
A_{2}=B_{1} $): Bob's and Alice's perspectives are formally
indistinguishable. In this case, no transition $B_{1}\rightarrow
A_{i}\rightarrow B_{2}$ or $B_{2}\rightarrow A_{i}\rightarrow B_{1}$ can
ever occur by probing the intermediate $\mathcal{A}$ perspective: letting
Bob probe Alice's perspective has no impact whatsoever on Bob's opinion
state ($2x\left( 1-x\right) =0$). Initial disagreement cannot be overcome
through deliberation.

ii. First note that from the point \ of view of the facilitator, the 2 two
consensus states are fully symmetric $\mathrm{prob}(B_{2}\rightarrow B_{1})=%
\mathrm{prob}(A_{1}\rightarrow A_{2})=2x\left( 1-x)\right) $ so a random
draw is optimal for the facilitator. The result in 1.ii. follows from Eq.~%
\eqref{Con2X2}, which shows that the probability to reach consensus is
strictly positive whenever $0<x<1$, namely when perspectives $\mathcal{A}$
and $\mathcal{B}$ are distinct.

iii. The probability for consensus in the first round
is\ maximal at $x=.5\ $where$\ 
\frac{\partial }{\partial x}2\left( 1-x\right) x\ =0,x\in \left[ 0,1\right]
.\ $The total probability for success after the second round, conditional on
failure in the first round, is $4x\left( 1-x\right) \left[ 1-2x\left(
1-x\right) \right] $ which reaches its maximum for uncorrelated perspective
as well i.e., $x=1/2$. For two rounds the maximum chance for consensus is $%
3/4$. 
\end{proof}
\medskip \ 

Result 1.i. is quite remarkable because it shows that sharing the same
thinking frame is an obstacle to achieving consensus when starting from
disagreement. Indeed, within a common thinking frame, citizens can only
update their opinion in response to new information (by Bayesian updating)
which we preclude in this paper. However, when citizens are endowed with
alternative perspectives new opportunities for opinion to evolve arise. By
actively exploring a perspective incompatible with ones' own, intrinsic
contextuality reveals its transformative power. Exploring an alternative
perspective changes the opinion state because the possible outcomes of that
operation do no exist in the own perspective. The opinion state is forced
into a new state. This result about the value of diversity is truly novel
and a main contribution of this paper. It is important at this point to recall \ref{assum_0}, the transformative value of
deliberation demands a true mental experience i.e., sincerely putting
oneself in someone else shoes - the probing operation.

Result 1.ii quantifies how the diversity of perspectives allows opinions to
evolve toward consensus. The weaker the correlation between perspectives$\
x\rightarrow .5$, the more impactful the probing operation. Starting from
disagreement $A_{1}B_{2},\ $the probability that Alice changes her opinion
from $A_{1}$to $A_{2}\ $is given by ${\rm Tr}\left( A_{1}B_{1}\right) {\rm Tr}\left(
B_{1}A_{2}\right) +{\rm Tr}\left( A_{1}B_{2}\right) {\rm Tr}\left( B_{2}A_{2}\right)
=2x\left( 1-x\right) $ which tends to zero as $x$ tends to 1(or 0) and it is
maximized at $x=.5$. The intuition is that the more closely related the perspectives the more
likely that probing Bob's frame takes Alice to the closest opinion state in
Bob's frame and when probing her own perspective anew she is most likely
confirmed in her initial opinion. Similarly for Bob, so disagreement is more
likely to persist. Nevertheless with some positive probability at least one
of the two citizens will end up having changed her/his mind which implies
consensus on voting. Interestingly, the result that uncorrelated
perspectives give the best chance for deliberation to achieve consensus,
reminds of a result in quantum persuasion (\cite{dan18b}). They show
that distraction understood as bringing attention to a perspective
uncorrelated to the targeted state is an efficient way to persuade Receiver.

Result iii. says, without surprise, that starting from dissensus additional
rounds following failure to reach agreement increase the probability for
consensus. While a single round can already achieve consensus with
probability $2\left( 1-x\right) x,$ with two rounds and uncorrelated
perspective case ($x=.5)$, we reach consensus in 75\% of the case. Of course
we do not expect citizens to repeat the same argument from round to round if
citizens are short of arguments that can be the time for experts with
suitable perspectives to be called in. \medskip\ \ 

\begin{corollary}
The first moving citizen has larger chance to see consensus on her initial
voting preferences than the one who moves second.
\end{corollary}

\begin{proof}
As earlier noted that the chance is the same whoever is selected: $%
prob\left( A_{1}B_{1};A_{1}B_{2}\right) =2x\left( 1-x\right) =prob\left(
A_{2}B_{2};A_{1}B_{2}\right) $. But since the second is only probed in case
of failure in the first round, the other consensual state has less chance to
be selected in the vote.
\end{proof}

The procedure gives more chance to the first selected citizen. The
introduction of a random draw restores the equality of chance between
citizens. \medskip

\begin{corollary}
When citizens's perspectives are correlated, relying on experts with
perspective uncorrelated to the current round's active citizen increases the
chance for reaching consensus in any given round.
\end{corollary}

This follows from result ii. When Alice and Bob have correlated perspectives,
the probability for reaching consensus when probing each other perspectives
is lower than 75\% after two rounds. The facilitator could choose instead
the following strategy. First, a random draw designates the active citizen. The procedure then proceeds as above except that only
experts are presenting arguments belonging to a perspective maximally uncorrelated
with the active citizen's.

Corollary 2 implies that there exists a tension between letting citizen
expose and probe each other's argument and the objective to maximize the score. This is not surprising given
result 1.ii. 
In order to preserve the democratic character of deliberation so it gives
voice to citizens, a mixture of citizen arguments and expert arguments can be
chosen at the cost of delaying consensus
however.

\subsubsection{Deliberations $n>2$}

We now consider the case when the perspectives have more than two dimensions
that is we have a finer characterization of the opinions. We remind that
probing a perspective corresponds to performing a complete measurement i.e.,
asking oneself which of the several possible (orthogonal) opinions you agree
with. The outcome of a probe is therefore always one of the possible
eigenstates of the probed perspective. Our main interest is for the
facilitator's choice of perspective to maximize the chance of consensus in
such a context. By Assumption \ref{assum_diff}, citizens' utility for the voting options
divides the $n-$dimensional state space into two subspaces. The Yes subspace
and the No subspace:%
\begin{eqnarray*}
Y^{j} &\equiv &\left\{ P_{k};\ EU^{j}\left( Y;P_{k}\right) \geq EU^{j}\left(
N;P_{k}\right) \right\} \\
N^{j} &\equiv &\left\{ P_{l};\ EU^{j}\left( N;P_{l}\right) >EU^{j}\left(
Y;P_{l}\right) \right\} ,j=A,B,\ k,l\in \left\{ 1,..,n\right\}
\end{eqnarray*}%
In constrast with the 2 dimensional case each citizens has a multiplicity of
opinion eigenstates which are consistent with the same voting option.

Consider a situation where Alice and Bob initially disagree on the voting,
say Bob votes Yes and Alice votes No. Assume that the random draw determines
consensus is first sought on Bob's position. So the first round of
deliberation is aimed at modifying Alice's voting intention from No to Yes.
Alice's opinion states $\mathcal{A}=(A_{1},\dots A_{n})$ are such that all $%
A_{i}$ are all pairwise orthogonal projectors in $\mathbb{H}^{n}$ (that is: $%
A_{i}A_{j}=0$ for all $i,j=1,\dots ,n$). We define the
subspace of Alice's opinion states leading to a Yes vote as $%
Y^{A}=(A_{1},\dots ,A_{k})$ and similarly for opinion states leading to a No
vote as $N^{A}=(A_{k+1},\dots ,A_{n})$. Without loss of generality, we
assume that Alice's initial opinion state is $A_{k+1}$. Hence, the aim of
the facilitator is to lead Alice to change her opinion state from $A_{k+1}$
(corresponding to an initial Yes vote), to any of the states $A_{i}$ with $%
i=1,\dots ,k$ (corresponding to a targeted No vote, in agreement with Bob's
vote).

We shall characterize the optimal strategy for the facilitator in terms of
the perspective that he invites Alice to probe. Typically, that will not be
Bob's perspective but a one that can be deployed by an expert\footnote{%
We could see this exercise as following two unsuccessful rounds where the
citizens probe each other's perspective. \ This is in order to preserve the
ideal that citizens' are invited to present their own argument to each other.%
}. Because the expert's perspective is distinct (and incompatible) from Bob's, Alice is
the only active citizen. Bob listens to the expert's argument but remains
passive (see Lemma 1 below). The general strategy of the facilitator is to
propose an (expert) perspective $\mathcal{C}=(C_{1},\dots ,C_{n})$,
incompatible with Alice's perspective (namely, $[A_{i},C_{j}]\neq 0$ for
some of the opinion states $A_{i}$ and $C_{j}$). As in the $n=2$ case, there
is a first transition $A_{k+1}\rightarrow C_{j}$ with probability $\mathrm{Tr%
}(A_{k+1}C_{j})$; then Alice updates her opinion leading to a second
transition $C_{j}\rightarrow A_{i}$ with probability $\mathrm{Tr}%
(C_{j}A_{i}) $. The probability for $A_{k+1}\rightarrow A_{i}$ is given by: 
\begin{equation}
\mathrm{prob}[A_{k+1}\rightarrow A_{i}]=\sum_{j=1}^{n}\mathrm{Tr}%
(A_{k+1}C_{j})\mathrm{Tr}(C_{j}A_{i})
\end{equation}%
where the sum is over the intermediate opinion states $C_{j}$ in the
perspective $\mathcal{C}$ proposed by (the expert selected by) the
facilitator.

We now investigate the optimal choice of perspective $\mathcal{C}$ in order
to maximize an opinion change $A_{k+1}\rightarrow A_{i}$ with $i=1,\dots ,k$
(namely, a change from an initial No vote to a target Yes vote). We below
show that the facilitator can achieve such a change with probability $p_{%
\mathrm{success}}=\frac{k}{k+1}$. This is obtained by choosing a perspective 
$\mathcal{C}=(C_{1},\dots C_{k+1},A_{k+2},\dots ,A_{n})$, with uniform
transition probabilities $\mathrm{Tr}(A_{k+1}C_{j})=\frac{1}{k+1}$ for all $%
j=1,\dots ,k+1$ (see Appendix \ref{app_building_C} for an illustration). In this case, we have: 
\begin{equation}
\mathrm{prob}[A_{k+1}\rightarrow A_{i}]=\sum_{j=1}^{k+1}\mathrm{Tr}%
(A_{k+1}C_{j})\mathrm{Tr}(C_{j}A_{i})+\sum_{j=k+2}^{n}\mathrm{Tr}%
(A_{k+1}A_{j})\mathrm{Tr}(A_{j}A_{i})
\end{equation}%
where the second sum vanishes thanks to the orthogonality of $A_{i}$
projectors: $\mathrm{Tr}(A_{k+1}A_{j})=0$ for all $j=k+2,\dots ,n$.
Furthermore, we notice that the transitions $A_{k+1}\rightarrow A_{i}$ with $%
i=k+2,\dots ,n$ are impossible, because in the first sum $\mathrm{Tr}%
(C_{j}A_{i})=0$ for $j=1,\dots ,k+1$ and $i=k+2,\dots ,n$ (this property is
a consequence of the context $\mathcal{C}$ being composed of pairwise
orthogonal projectors). Only transitions $A_{k+1}\rightarrow A_{i}$
with $i=1,\dots ,k+1$ are possible. Hence, Alice maintains her
initial voting intention only when moving back to $A_{k+1}$ which happens with probability:%
\begin{eqnarray}
\mathrm{prob}[A_{k+1} &\rightarrow &A_{k+1}]=\sum_{j=1}^{k+1}\mathrm{Tr}%
(A_{k+1}C_{j})\mathrm{Tr}(C_{j}A_{k+1})  \label{eq_p_failure_1} \\
&=&\sum_{j=1}^{k+1}\left( \frac{1}{k+1}\right) ^{2}=\frac{1}{k+1}  \notag
\end{eqnarray}%
In this case, the facilitator fails to reach an agreement between Alice and
Bob. In all other cases, Alice's new opinion state $A_{i}$ with $i=1,\dots
,k $ leads to a Yes vote, hence consensus is achieved, and this happens with
probability $p_{\mathrm{succes}}=1-1/(k+1)=k/(k+1)$.

Proposition \ref{prop_2} below collects these insights

\begin{proposition}
\label{prop_2}
When citizens' perspective are $n-$dimensional and the dimensionality of the
subspace of the projected consensus for the active citizen is $k$,

i. consensus can be reached a success probability equal $p_{\mathrm{success}%
}=k/(k+1);$

ii. the perspective $\mathcal{C}$ that delivers $p_{\mathrm{success}%
}=k/(k+1)\ $is of the form $\mathcal{C}=(C_{1},\dots ,C_{k+1},A_{k+2},\dots
A_{n})$,with $\mathrm{Tr}(A_{k+1}C_{j})=1/(k+1)$ for all $j=1,\dots ,k+1$.

iii. $C$ is optimal within a natural class of perspectives.
\end{proposition}

\begin{proof}
see Appendix \ref{app_proof_P2} \medskip
\end{proof}

Proposition 2.i and 2.ii establish that our results from Proposition 1
generalize to multidimensional perspectives. Indeed letting $k=1,$ we
recover our previous result namely that $x=\mathrm{Tr}%
(A_{k+1}C_{j})=1/(k+1)=1/2$ that is uniformally uncorrelated perspectives
maximize the chance of consensus$.\ $We also find that the multiplicity of
dimensions increases the chance of consensus when properly addressed. The
probability of success is proportional to the dimensionality of the subspace
associated with the projected consensus vote, $k/(k+1)$. So it is easier for
people to reach agreement the finer their initial (own) perspective and the
larger the set of opinion states consistent with the projected consensus
vote (here Yes).

A new result in 2.ii is that the optimal perspective is characterized by
effectively reducing the dimensionality of the problem. This is done by
selecting $C$ such that $C_{k+1}$ to $C_{n}$ are identical to $A_{k+1}$ to $%
{A}_{n}$\ implying that there is no chance that deliberation results in
a new opinion state that yields a No vote except for moving back to the
initial state $A_{k+1}$. We can say that the dimensionality (in the probing
operations) is effectively reduced from $n$ to $n-k\ .$\ This reminds of an
often heard argument saying that deliberation should aim at reducing the
scope of disagreement.

To conclude that this strategy is globally optimal, one must show that other
structures of the perspective $\mathcal{C}$ namely those allowing for
transitions $A_{k+1}\rightarrow A_{i}$ with $i=k+2,\dots ,n$ cannot increase
the probability to achieve a transition $A_{k+1}\rightarrow A_{i}$ with $%
i=1,\dots ,k$. This conjecture is intuitive, but remains to be formally
proved.

\begin{corollary}
To maximize the probability for consensus, the facilitator should ask the
citizen who has the largest chance to change her voting intention to be the
active one. This is the citizen whose alternative to her current voting
intention is supported by a larger set of opinion states than the corresponding set for the
other citizen.\medskip\ 
\end{corollary}

This means that in a multidimensional case, the two possible consensus
states are not symmetric which contrast with the 2-dimensional case. In
particular, if $\dim Y^{Alice}<\dim N^{Bob},$ it is easier to achieve
consensus by having Bob change opinion from Yes to No than Alice from No to
Yes: $\frac{\partial }{\partial k}\left( \frac{k}{\left( k+1\right) }\right)
>0$. We shall see below that asymmetry between voting options in terms of
ease to reach consensus arise for different reasons. Therefore, we introduce
a new feature that we call \textit{projected consensus state} corresponding
to the consensus state that the facilitator projects to reach.

\textbf{Definition}

We refer to as \textit{projected consensus state,} the consensus state that
is targeted by the facilitator.\medskip\ 

Finally, One may want to question the practical value of the result in
Proposition 2. First, it is of course very unlikely that a citizen's
perspective exhibits the desired properties of the optimal perspective.
Therefore, the facilitator should consider turning to an expert - possibly
after two unsuccessful rounds of reciprocal probing between Alice and Bob.
Constructing the optimal $C$ which is a new perspective implies the
"relabelling" of some opinion states from another perspective. Basically
when $C_{1}=A_{1}$ it means that the two opinions while being represented by
the same state are expressed "in two different languages"\footnote{%
For instance, if $A_{1}$ is opinion "ICB is the environmentally best way to
reduce GHG", $C_{1}$ could be "providing proper incentives to individuals is
the most efficient way to reduce GHG".}.

\subsection{Deliberation in a population of citizens $\left\vert \Omega
\right\vert >2$}

We next analyze deliberation in a population of voters like a citizen
assembly. We consider the simpler but most relevant case when the population is divided into two groups each with its own frame
of the voting issue. The ICB example is a suitable one as it relates to
quite well establish ideologies a left leaning social and environmental
ideology ($L)\ $and a right leaning conservative libertarian ideology ($R)$.

We first establish a simple Lemma

\begin{lemma}
Assume that two citizens with
alternative perspectives agree on voting, inviting any of them to probe the other's perspective jeopardizes consensus. 
\end{lemma}

\begin{proof}
Consider starting from $(L_{1},R_{1})$ and letting the ${\cal R}$-citizen probes the ${\cal L}$
perspective, we are back in $(L_{1},R_{1})$ with probability $%
{\rm Tr}(R_{1}L_{1}){\rm Tr}\left( L_{1}R_{1}\right) +{\rm Tr}(R_{1}L_{2}){\rm Tr}\left(
L_{2}R_{1}\right) = x^{2}+(1-x)^{2}<1$, so consensus is lost with
positive probability for $x\notin\{0,1\}$ that is when the two perspective are distinct from each other. 
\end{proof}

Lemma 1 shows that having citizens explore alternative perspectives can be
harmful to existing agreement among citizens. This result underscores the central role of the facilitator in our context. There is no necessity that deliberations lead to consensus. Instead, the path of opinion change depends of the process itself; more precisely it hinges on which citizens are active and which perspective they probe in each round. We note that Lemma 1 is
consistent with much criticism appealing to cognitive biases that emphasizes
that interacting politically can have "negative" impact on citizens' beliefs
and preferences \cite{janis82,Elster05}. \medskip

Let the two groups be of size$\ l$ respectively $r$. Typically the population is characterized by disagreement both within and between the two groups. In
each group the citizens have probed their own perspective, so we have two distributions $\left(l_{1},l_{2}\right) ,\ l_{1}+l_{2}=l$ and similarly $\left(
r_{1},r_{2}\right) ,\ r_{1}+r_{2}=r$. Let the two consensual states be $(L_{1},R_{1})$ and $(L_{2},R_{2}$. In each round the facilitator seeks to maximize the expected score. Consider the
case when citizens (rather than experts) present their own arguments, we have:
\begin{proposition}
Starting from disagreement within and/or between 2 groups, in any round\newline 
i. the projected consensus need not be the standing majority;\newline
ii. the facilitator invites citizens disagreeing with the projected consensus to explore the other perspective, while the remaining citizens remain passive until next round.
\label{Prop_3}
\end{proposition}

\begin{proof}
3.i. We know from Proposition 1 that the probability for opinion change is $2x(1-x)=\Delta$. The expected change is thus $\Delta l_{i}$ (or $\Delta r_{j}$) where $i,j=1,2$ depending on the group invited for probing. Consider a case where $(l_{1}+r_{1})>(l_{2}+r_{2})$, so the majority supports Yes. Then whenever $l_{1}+r_{1}+\Delta \max(l_{2}, r_2) < (l_{2}+r_{2})+\Delta \max(l_{1}, r_1)$, that is $(l_{2}+r_{2})-(l_{1}+r_{1})>\Delta[\max(l_{2},r_2)-\max(l_{1}, r_1)]$, the optimal projected state corresponds to the No (minority) vote. When the inequality goes the other way the standing majority is the optimal projected consensus state.  3.ii Given 3.i the largest group of citizens disagreeing with the projected consensus is identified. Because of Lemma 1, no other citizen from that perspective group is invited to probe. 
\end{proof}

Clearcut quantitative predictions echoing our results in \ref{Prop 1} can be formulated:
\begin{corollary}
 With maximally uncorrelated perspectives, starting with two perspective groups of the same size and with evenly distributed voting intentions, deliberation delivers a (expected) majority of 3/4 of the population after two rounds. 
   
\end{corollary}
\begin{proof}
    Let $l_{1}=l_{2}=l/2$ and similarly $r_{1}=r_{2}=l/2$ with $\Delta=1/2$. Any of the two consensus states can be reached with the support of a population of size $l/2+r/2+l/4+r/4=(3/4)(l+r)$.
\end{proof} 
  \medskip

The results in Proposition \ref{Prop_3} invite multiple remarks. 
First, we learn that the projected consensus state need not be the standing majority. It is quite a remarkable result because it applies when the facilitator is ``myopic" i.e., maximizes the score at each step. We view this as a nice property of our deliberation procedure, in the sense that it gives some chance to the minority voting option. We return below to other circumstances where the minority can be given the chance. 

 A second important remark is that the optimal strategy relies on the distinction between active and passive citizens. Because of Lemma 1, when both opinions are present in a perspective group, it is optimal for the facilitator to proceed selectively. Those from the targeted perspective group already agreeing with the projected consensus state should refrain from probing, as they could change their mind and reduce the score. This means that the optimal procedure demands some extent of differentiation between citizens, which is unfortunate from a democratic ideal point of view. Note however that full publicity of debates can be preserved because with contextual opinions, only the operation of probing can induce change. In our model, simply listening to an argument without making the effort of thinking in the terms of the alternative perspective has no effect on opinions.

Our analysis of the population case provides a novel rationale (and guiding principles) for the practice of parallel working groups encountered in real life deliberations.\footnote{Obviously, the practice also speeds up the process which is presumably a main motivation.} Our approach suggests that this practice also responds to the risks that deliberation ``breeds confusion" in people's mind. With well-composed parallel working groups, one can prevent unwanted opinion switch when citizens make probing operations without being invited to.

\subsection{Deliberation with more than two competing perspectives}
\label{sec_3P}

Let us now consider the case with citizens representing more than two
competing perspectives. Assume three citizens are endowed each with their own
two-dimensional thinking frame. This context will also allow addressing the
restriction on the facilitator's strategies that asks him to maximize the
score in each round.

Continuing on our ICB example, we now have Greg with a perspective concerned
with legal/practical feasibility of ICB and two eigenstates e.g., $G_{1}:\ $%
ICB can be integrated in the current set of laws and regulation with
reasonable adaptation costs; $G_{2}$ enforcing ICB requires major costly
legal adaptation. \medskip

As in the population case, since the vote is binary, we initially always have at least
two citizens who agree on voting. Hence the two consensus states, let them
be $(A_{1}B_{1}G_{1})$\ and $(A_{2}B_{2}G_{2}),$ are not symmetric$\ $with
respect to the initial state. This feature was also encountered in the 
$n-$dimensional and population case.\medskip\ \ 

\subsubsection{Maximizing the probability for consensus in each round}

We first investigate the facilitator's strategy when restricted as before to
maximizing the score in each round -- we thereafter relax that constraint and
consider maximization over two rounds. In this context, it seems fair and common sense to demand that the same perspective not be presented twice. By Proposition 3, we expect the facilitator to select the majority
voting option as the projected consensus state. If the initial opinion state is $(A_{1}B_{2}G_{1}),\ $the natural
candidate strategy entails focusing on the sole disagreeing citizen, Bob. As we focus on the voice phase where citizens expose their own views and appealing to Lemma 1, the disagreeing
citizen (Bob) is the sole active citizen, Alice and Greg should never take 
\textit{active} part in the deliberations. We have the following result:

\begin{proposition}
\label{prop4}
Assume that disagreement with respect to voting characterizes a group of
three citizens belonging to three distinct perspectives. Consensus can be approached
with selective targeted deliberation. It entails i. selecting the majority
voting option as the projected consensus state in ii. targeted deliberation
where only the disagreeing citizen is active in all rounds. To maximize the probability of consensus in the first round, the disagreeing citizen should be first exposed to the least correlated perspective. Yet, the order in which the two perspectives of the majority citizens are exposed to the disagreeing citizen is irrelevant to the overall probability of consensus.
\end{proposition}

\begin{proof}
    See appendix \ref{app_proof_p4}.
\end{proof}

The results in Proposition \ref{prop4} invites two remarks. First, we find that
the presence of more than two perspectives in
a population of citizens does not change the facilitator's problem significantly. He optimally orders the perspectives that he invites Bob to probe so as to start with the least correlated. So the multiplicity of perspectives neither facilitates the task of the facilitator nor is it an obstacle. As in Proposition 3, the deliberative process is skewed towards the opinion which is
initially majoritarian. The process essentially amounts to persuading Bob to change
opinion while Alice and Greg listen to new arguments but refrain from
exploring any new perspective. Result 4.ii contradicts the principle of equity as Bob is never offered the opportunity to
convince Alice and Greg to change their initial opinion, and reach the
alternative consensus state $A_{2}B_{2}G_{2}$. We next consider relaxing the
restriction on the facilitator. If his task is not to maximize the
probability of consensus \textit{in each round} but over a fixed number of rounds, may he choose to
upset the standing majority? opening the way to reach the alternative
consensus state $A_{2}B_{2}G_{2}$. We next consider the two-rounds case\medskip 

\subsubsection{Maximizing the consensus probability over two rounds}
As in the previous subsection, we let the
initial vector of opinion state be ($A_{1}B_{2}G_{1}$). We consider an alternative strategy
for the facilitator, where in the first round two citizens are active e.g., Bob and Greg who
are invited to probe Alice's perspective. There are four possible outcomes of
this first round: $%
(A_{1}B_{1}G_{1}),(A_{1}B_{2}G_{1}),(A_{1}B_{2}G_{2}),(A_{1}B_{1}G_{2})$.\
This implies that in the second round, where the facilitator maximizes the immediate probability for consensus, the two consensus states $%
(A_{1}B_{1}G_{1})$ and $(A_{2}B_{2}G_{2})$ may emerge. We have the
following Proposition

\begin{proposition}
\label{prop5}
Assume that disagreement with respect to voting characterizes a group of
three citizens belonging to three distinct perspectives. When the facilitator is
aiming at reaching consensus within a 2-period time frame using the citizens'
perspective exclusively, shortsighted maximization is not always optimal.
Instead, jeopardizing the standing majority may give higher chance for
consensus and anyone of the two voting options may prevail. 
\end{proposition}

\begin{proof}
See Appendix \ref{app_3citizens}.
\end{proof}

In Appendix \ref{app_3citizens} we characterize the conditions under which letting Bob
challenge both Alice and Greg can generate a better outcome than focusing on
persuading Bob. The conditions corresponds to a case where Bob is closer
(his perspective is more correlated) to both Alice and Greg than Greg is to
Alice. This is illustrated with a numerical example 

\begin{figure}
    \centering
    \includegraphics[width=0.6\linewidth]{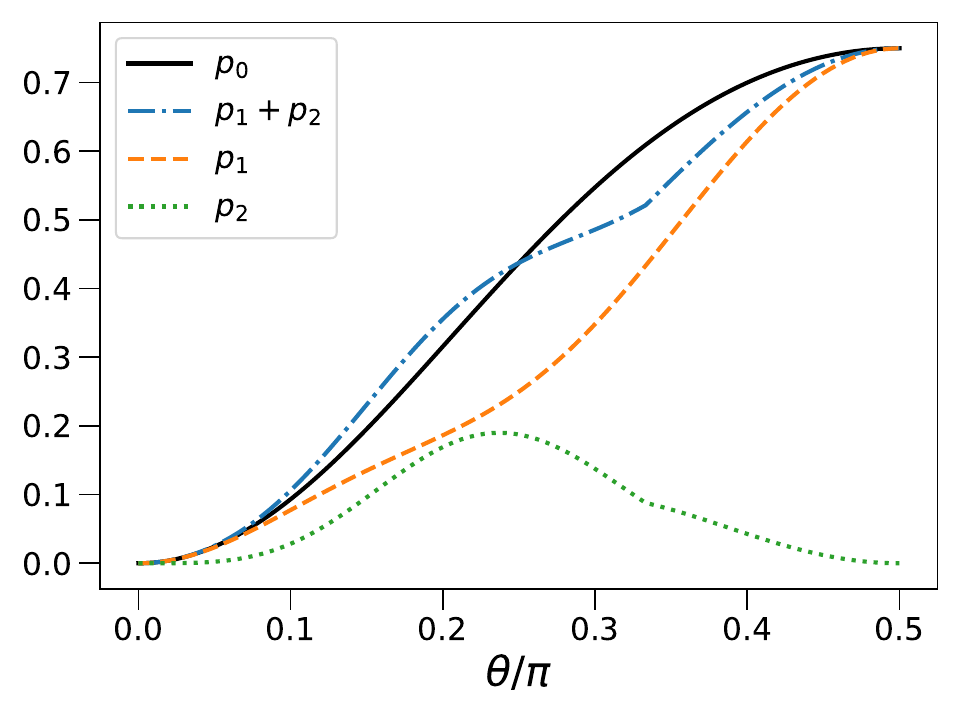}
    \caption{\scriptsize Two-rounds deliberation with three citizens in the two-dimensional case, starting from state $(A_1 B_2 G_1)$. $p_i$ are probabilities to reach consensus states after two rounds in different scenarios. See text for a definition of $p_0, p_1, p_2$, and Appendix \ref{app_3citizens} for their relation to $\theta$.}
    \label{fig:numerical_illustration}
\end{figure}

\textit{Numerical illustration.--} For a concrete characterization of this
situation in the quantum formalism, we introduce the angle $\theta $ such
that $\mathrm{Tr}(A_{1}B_{1})=x=\cos ^{2}(\theta /2)$. We also assume that $\mathrm{Tr}(B_{1}G_{1})=x$; and we assume that $\mathrm{Tr}(A_{1}G_{1})=y=\cos
^{2}(\theta )$, namely, the unit vector representing $A_{1}$ opinion state
forms an angle $\theta /2$ with the vector representing $B_{1}$ opinion
state, which forms itself an angle $\theta $ with the vector representing
the $G_{1}$ opinion state. The whole correlations between Alice's, Bob's and Greg's perspectives are therefore entirely defined via a single parameter $\theta$, which allows for a simplified illustration of the different possibilities. We focus on the parameter regime $\theta \in
\lbrack 0,\pi /2]$, relevant to describe the two cases described above. We introduce $p_0$ as the probability to reach consensus $(A_1 B_1 G_1)$ by letting Bob probe first Alice's and then Greg's perspective. $p_1$ is the probability to reach consensus $(A_1 B_1 G_1)$ if both Bob and Greg actively probe Alice's perspective in the first round; and $p_2$ is the probability to reach consensus $(A_2 B_2 G_2)$ in this same case. The dependence with $\theta$ of $p_0$, $p_1$, $p_2$ and $p_1+p_2$ is plotted on Fig.~\ref{fig:numerical_illustration}.

Our main finding is that for $\theta \in [0,\pi /4]$, we have $p_1+p_2 \ge p_0$, namely it is better to let both Bob and Greg be active during the first round where Alice exposes her perspective. In this scenario, the majority (Alice and Greg) can been
converted to the minority's initial opinion (Bob's) through the deliberation
procedure with a sizable probability $p_2$ (green-dotted line in Fig.~\ref{fig:numerical_illustration}); and this allows to reach consensus
with a higher success rate ($p_1+p_2$, blue dashed-dotted line) than insisting twice in converting the minority
opinion (in our case, Bob) to the majority view (Alice and Greg) ($p_0$, black-solid line). On the
other hand, for $\theta \in \lbrack \pi /4,\pi /2]$, the facilitator
maximizes the consensus probability when trying twice to reverse Bob's initial opinion, letting Greg and Alice play a less engaged role.

Proposition \ref{prop5} shows that it can, under some conditions, be
optimal to allow for disrupting the standing majority and give chance to the other voting
option. As shown in the example (and in the proof of Proposition 5) this
happens when the disagreeing citizen's perspective is more closely
correlated to both of the two agreeing citizens' perspective than the
agreeing citizens' perspectives among themselves. We notice however that the most probable consensus state remains the one consistent with the initial majority (namely $p_1 \ge p_2$, respectively dashed-orange and green-dotted lines in Fig.~\ref{fig:numerical_illustration}).

\section{Discussion}

\subsection{Learning under complete information}

In this section, we wish to argue that while we developed the whole analysis
in a context of complete information so that at each moment the citizens are
endowed with a complete characterization of the issue at stake, some form of
learning does nevertheless take place.

The central hypothesis of this paper, is that there exist perspectives that
people cannot consider simultaneously, they are Bohr complementary. While
incompatible perspectives can be considered in sequence, people cannot
aggregate information from two incompatible frames \textit{in a stable way}.
This instability is the expression of intrinsic contextuality meaning that
opinions do not "exist" (have a determinate value) independently of a
thinking frame (perspective) and that it is impossible to integrate all
frames into a unique "super frame". Instead, there exists a multiplicity of
alternative frames that are equally valid to \textit{fully }characterize an
issue.

Nevertheless when probing an alternative perspective, the citizen learns
something because the new state corresponding one of the eigenvalues of the
perspective under exploration does not exist (as a pure state) in the
initial perspective. An alternative frame is like a new language that can
express something that could not be expressed in the original language
(initial frame)\footnote{%
Of course any state can be expressed in any perspective but except for the
eigenstate, they will be mixtures of other (eigen)states.}. The exploration
of an alternative frame comes at the cost of previously held information:
the opinion state is modified by the probing operation. This is why we can
speak about learning without learning \textit{more}. We believe that our
application to deliberation sheds a new light on the concept of "Bohr
complementarity", more specifically on the \textit{complementarity} of
information from incompatible perspective.\ Niemeyer et al (\cite{niemeyer07}%
) speak about "Intersubjective representational framework" and "meta
representation". Our approach allows capturing the meta representation\ as
knowledge of the multiplicity of valid ways to structure one and the same
issue. Physicist and philosopher, Michel Bitbol talks about the "second
order objectivisation"\footnote{%
That is a procedure of objectivisation of the methodes of production and
anticipation of phenomena that cannot be objectivised as properties
belonging to objects. .} \cite{bitbol09}. The meta-representation which
emerges from our deliberation process corresponds to the state space
(representing the issue) and the set of associated bases representing
different incompatible perspectives together with the correlation between
them. So citizens do learn something in deliberation namely how a problem
can be considered in ways different from their own (alternative
perspectives) and how these ways related to their own (correlation between
perspectives) that is they learn elements of the "super structure" of the
opinion space. We believe that apart from having one's own opinion evolving,
this kind of more fundamental knowledge contributes to making deliberation a
process that transform people as often reported. This also illustrates what
some participatory democrats with John Stuart Mills (see Introduction) have in mind when assessing that deliberative democracy has a
value in itself. It educates people as citizens of a democratic society.
Learning the super structure of the opinion space from experiencing (and
thus recognizing the legitimacy) a variety of perspectives, can be understood in this light.

This does not affect the epistemic value of the decision finally made,
however. Our approach cannot provide an answer to that question. In the
profession, beyond standard informational arguments, the question remains by
large open (see Estlund and Landemore in \cite{estlundlande18}). With our
approach, deliberation actualizes one of the deepest debates in QM (see \cite%
{despa12}). Indeed even at the individual level, the "correctness" of a
decision is contextual. What is the right decision from one perspective need
not be the right decision for the same individual from another perspective,
thus allowing deliberations to lead to changes in preferences. Therefore,
the common aggregative argument has no bite on the epistemic value of the
resulting decision. It is beyond the scope of this paper to formulate a
precise relationship between the epistemic value of deliberations i.e.,
citizens learn about the super structure of the opinion space, and the
epistemic value of the decision. We conjecture that a result in the spirit
to L. Hong and S. Page "Diversity trumps ability" \cite{hongpa04} is a
promising path to go in a (quantum) non-classical context.

\subsection{The deliberative democratic ideal}

Our approach has been successful in characterizing the transformative
character of deliberations: opinions evolve when exploring alternative
thinking frames without improvement in information. The analysis shows
not surprisingly that deliberations are subject to a tension between
efficiency in terms of the probability to reach consensus and fairness in
terms of equal treatment of the citizens. Traditionally, equality in the
context of deliberation refers either the problem of including all concerned citizens \cite{dryzek01} or of redressing situations of inequality in
terms to social cultural resources that affect their ability to participate
on equal footing in deliberative exchanges (see \cite{beauvais18}). This is
not our concerns here. We are interested in inequalities that arise because
convergence to consensus may require breaking some symmetry in the treatment of participating  citizens.

The approach allows characterizing two main aspects of this inequality: the
asymmetry between voting options determined by the initial state of opinions
and the distinction between active and passive citizens in the process of
deliberation. Regarding voting options, we establish in Proposition \ref{Prop_3} and \ref{prop5} that it may be optimal to give a chance to the standing minority option. In general however, the initial majority option tends to be determinant for the final outcome.   As suggested in the text, the procedure could start with a "voice phase" where citizens are invited to probe each other's
perspectives with no immediate demand to maximize consensus. This phase would determine an initial majority option.\footnote{Additional analysis is needed to determine the structure of the voice phase.} It
would be followed by an "expert-led phase" where the facilitator maximizes the
support of the most supported option.  The latter phase which actualizes the
distinction between active and passive citizens implies differentiated treatment. Only
minority voters are subjected to persuasion. In defense of
this unequal treatment, we wish to emphasize that it involves transparent and public
operations of (quantum) persuasion. In addition, the participants would be
invited to agree with the terms of the procedures before entering the whole
process i.e., before knowing in which role they might be\footnote{%
The alternative to deliberation for those who reject its terms would be to
participate to the vote directly.}. Although not ideal, we would like to
argue that with a benevolent facilitator, the optimal procedures that we
have analyzed are not irreconcilable with the ideal of democratic fairness.

\textit{Caveats 1}

We have addressed deliberations in terms of the standard quantum formalism.
As well-known this formalism entails quite a lot of formal constraints which
may not all be meaningful in the context of social sciences. A more general
framework e.g., appealing to POVM (positive operator-valued measure) could
be more appropriate. We therefore propose the current analysis as an
abstract approximation which we believe provides interesting insights and
can be submitted to experimental testing.

\textit{Caveats 2}

In our model the facilitator is fully benevolent. In view of his influence
on the outcome, an important issue to investigate is how to secure proper
incentives. We leave this important question for future research.

\section{Conclusions}

In this paper we have developed a formal approach to deliberation based on
the behavioral premises and mathematical formalism of quantum cognition. The
behavioral premises include i. people need a thinking frame (perspectives)
to address reality, ii. Citizens cannot (always) combine all relevant
aspects in a single thinking frame; iii. alternative thinking frames are
Bohr complementary: they are both incompatible and equally valid
perspectives on the issue at stake.

Deliberation is formulated in the context of complete information as a
structured communication process managed by a facilitator with the aim of
maximizing the probability for consensus in a binary collective choice problem. The process
includes a sequence of rounds in which some citizen or experts develop an
argument belonging to some perspective and other citizens are invited by the
facilitator to probe that perspective. Probing involves a true action from
citizens, they actively "put themselves in someone else's shoes" and decide
how they position themselves in a perspective alternative to their own.

A first central result is that the incompatibility of perspectives, that is
the diversity of view points between disagreeing citizens is what permits
opinions to evolve. Our second central result is that the correlation
between perspectives is the key property that determines the pace of
evolution toward consensus. In the two citizens, two perspectives and two
dimensional case, starting from disagreement, consensus is reachable with
75\% chance after two rounds only. This result generalizes to the case when
perspectives have more than 2 dimensions. In that context, consensus is
achievable after one round with a probability that can approach 1. The highest
rate of convergence to consensus is achieved when the citizen's initial
perspective and the one she is invited to probe are maximally uncorrelated that
is when having an opinion in one perspective give equal chance for any
opinion in the other perspective. In the multidimensional case, the strategy
of the facilitator also involves reducing the dimensionality of the problem.

The results generalize to more than 2 citizens where the facilitator's
strategy involves selective targeting in the sense that while deliberation
remain fully public only a selected subset is invited to actively probe the
presented perspective in each round. The others simply listen. In the larger population context and in the presence of multiple perspectives, the facilitator may choose to challenge the standing majority in particular when he does not maximize the score myopically. 

We thus find that the quantum cognition approach allows giving sense to a
number of empirical features put forward in the literature. Most
importantly, it delivers the transformative character of deliberation which
goes far beyond Bayesian updating. In our model, people go through real
(mental) experiences (probing) which transform their opinion and deepen
their understanding of each others. While in the model they always are in a
situation of complete information, they learn how the issue at stake
can be approached from equally valid alternative perspectives and how these
relate to their own. Our analysis is also consistent with the proposition often put
forward that deliberation requires serious engagement and fosters the respect for
each others as a practical school of democracy. In our context this is
captured by the "effort" of "putting oneself in someone else's shoes" which implies the respect of other citizens' perspective. Finally, Our
approach characterizes the determinant role of the facilitator(s) who
plays a central role in most actual experiments such as the "Convention
Citoyenne pour le climat" (2021) in France. Finally and importantly, it offers an
approach to reaching consensus in deliberation not based on improved information.

\pagebreak

\pagebreak

\appendix

\section{Building an intermediate perspective}
\label{app_building_C}
We illustrate here that it is indeed possible to build such a perspective $%
\mathcal{C}.$ Formally, a change of perspective is given by a $n\times n$
unitary matrix $U$, representing a change of orthonormal basis. The opinion
states are then the projectors onto the corresponding basis states. If $%
|a_{i}\rangle $ are the basis vectors for Alice's own perspective, and $%
|c_{j}\rangle $ are the basis vectors for the intermediate perspective
proposed by the facilitator, the property $\mathrm{Tr}(A_{k+1}C_{j})=|%
\langle a_{k+1}|c_{j}\rangle |^{2}=1/(k+1)$ for all $j=1,\dots ,k$ is for
instance obtained if 
\begin{equation}
|a_{k+1}\rangle =\frac{1}{\sqrt{k+1}}\sum_{j=1}^{k+1}|c_{j}\rangle ~.
\end{equation}

\bigskip

\section{Proof of proposition \ref{prop_2}}
\label{app_proof_P2}
Given perspective ${\cal C}$, the probability of failure is simply given by the probability
that Alice maintains her initial opinion state, namely to make the
transition $A_{k+1}\rightarrow A_{k+1}$. According to Eq.~%
\eqref{eq_p_failure_1}, this occurs with probability $p_{\mathrm{failure}%
}=\sum_{j=1}^{k+1}w_{j}^{2}$ where we introduced the notation $w_{j}=\mathrm{%
Tr}(A_{k+1}C_{j})$, which are such that $\sum_{j=1}^{k+1}w_{j}=1$. Hence, we
have to optimize the $w_{j}$ such as to minimize $p_{\mathrm{failure}}$,
under the constraint $\sum_{j}w_{j}=1$. Introduce a Lagrange multiplier $%
\lambda $ to enforce the constraint, and introduce the function $%
f(w_{1},\dots ,w_{k+1})=\sum_{j}w_{j}^{2}-\lambda (\sum_{j}w_{j}-1)$. The
minimum of $f$ is achieved when $%
\frac{\partial f}{\partial w_{j}}=0$ for all $j$, namely when $w_{j}=\lambda
/2$ for all $j$. That is: the minimum is achieved for a uniform choice of
the $w_{j}$. Since we must have $\sum_{j=1}^{k+1}w_{j}=1$, this leads to the
result $w_{j}=1/(k+1)$, and to the minimal failure probability $p_{\mathrm{%
failure}}\geq 1/(k+1)$; and correspondingly: $p_{\mathrm{success}}\leq
k/(k+1)$.

\section{Proof of Proposition \ref{prop4}} 
\label{app_proof_p4}
Consider Bob disagreeing because he holds opinion $B_{2}$ while Greg holds
opinion $G_{1}$ and Alice, $A_{1},$ they both vote Yes$.$\ Let the
facilitator selects $A_{1}B_{1}G_{1}$ as the projected consensus state. The
total probability for success is $\mathrm{Prob}%
^{t=1}(B_{2}\rightarrow B_{1})+[1-\mathrm{Prob}^{t=1}(B_{2}\rightarrow B_{1})]\mathrm{Prob}^{t=2}(B_{2}\rightarrow B_{1})$. In the first round he chooses among Greg and Alice. Assume $\mathrm{Tr}\left(
B_{1}A_{1}\right)>\mathrm{Tr}\left(
B_{1}G_{1}\right)$, then
he let's Alice develop her argument for Bob so he probes her perspective.
Global consensus $A_{1}B_{1}G_{1}$ is reached if
Bob's opinion state changes from $B_{1}$ to $B_{2}$. This happens with
probability $\mathrm{Prob}^{t=1}(B_{2}\rightarrow B_{1})=\mathrm{Tr}\left(
B_{2}A_{1}\right) \mathrm{Tr}\left( A_{1}B_{1}\right) +\mathrm{Tr}\left(
B_{1}A_{2}\right) \mathrm{Tr}\left( A_{2}B_{1}\right) =2x(1-x)=a$, with $x=%
\mathrm{Tr}(A_{1}B_{1})$. In case consensus fails, the same reasoning
applies for the second round: Greg is now invited to present his perspective
to Bob, who changes opinion with probability $\mathrm{Prob}%
^{t=2}(B_{2}\rightarrow B_{1})=\mathrm{Tr}\left( B_{2}G_{1}\right) \mathrm{Tr}%
\left( G_{1}B_{1}\right) +\mathrm{Tr}\left( B_{1}G_{2}\right) \mathrm{Tr}%
\left( G_{2}B_{1}\right) =2z(1-z)=b$, with $z=\mathrm{Tr}(G_{1}B_{1})$. The
total probability for success is $\mathrm{Prob}%
^{t=1}(B_{2}\rightarrow B_{1})+(1-\mathrm{Prob}^{t=1}(B_{2}\rightarrow B_{1})\mathrm{Prob}^{t=2}(B_{2}\rightarrow B_{1})=a + (1-a)b$. If instead Greg exposes first his perspective, and then Alice in case of failure, the total probability of success is $b + (1-b)a = b + a - ab$, namely the same result. Hence, the order in which Alice and Greg expose their perspectives to Bob is irrelevant to the overall probability of success.

\section{Proof of Proposition \ref{prop5}}
\label{app_3citizens}
We define $a=2x(1-x)$, $b=2y(1-y)$ and $a^{\prime }=2z(1-z)$ with $x={\rm Tr}(A_1 B_1)$, $y={\rm Tr}(A_1 G_1)$ and $z=\mathrm{Tr}%
(G_{1}B_{1})$. $a$ is the probability that Alice (resp. Bob) changes opinion when exposed to Bob's (resp. Alice's) perspective; $b$ is the
probability that Greg (resp. Alice's) changes opinion when exposed to Alice's
(resp. Greg's) perspective; etc. Recall that the initial state is $(A_1 B_2 G_1)$. We consider that Alice's perspective is exposed first, and compare the case where: 1) only Bob is active in the first round; and 2) both Bob and Greg are active in the first round.\\

\textit{Only Bob is active in the first round.--}
In this scenario, the probability to reach
consensus (consensus state $A_{1}B_{1}G_{1}$, namely Bob has change his
opinion to $B_{1}$) at the first round is $a$. If his opinion remains $B_2$, then Bob is exposed to Greg's perspective and changes opinion with probability $a'$. Overall, the probability to reach consensus over two rounds is $p_{0}=a+(1-a)a^{\prime }$, where the
first term ($a$) is the probability that Bob changes opinion at the first
round, and the second term ($(1-a)a^{\prime }$) is the probability that Bob
changes opinion at the the second round, conditioned on the fact that he did
not change opinion at the first round.\\

\textit{Both Bob and Greg are active in the first round.--}
The probability to reach consensus in the first
round is $a(1-b)$, which is obviously smaller than in the previous scenario,
as there is now a non-zero probability that Greg also changes opinion,
reaching the state $(A_{1}B_{1}G_{2})$ instead of $(A_{1}B_{1}G_{1})$. But
as we shall see, this may allow for a higher overall probability to reach
consensus over the two rounds. If $(A_{1}B_{1}G_{1})$ is not reached at the
first round, then the facilitator may choose to present either Bob's or
Greg's perspective; and lets the minority citizen probe that perspective,
offering her/him to change opinion. Let us enumerate the possibilities:

\begin{enumerate}
\item If the state after the first round is $(A_{1}B_{2}G_{1})$, which
happens with probability $(1-a)(1-b)$, then Greg presents his perspective to
Bob, while Alice remains passive: consensus $(A_{1}B_{1}B_{1})$ is then
reached with probability $a^{\prime }$. The overall probability to follow
this scenario is hence $(1-a)(1-b)a^{\prime }$. 

\item If the state after the first round is $(A_{1}B_{2}G_{2})$, which
happens with probability $(1-a)b$, then the targeted consensus state is $%
(A_{2}B_{2}G_{2})$: Alice is now the active citizen, and she is offered to
probe either Bob's or Greg's perspective. She changes opinion with
probability, respectively, $a$ or $b$. Then facilitator hence chooses the
perspective which maximizes this probability. Hence, overall, the
probability to follow this scenario and end up in the $(A_{2}B_{2}G_{2})$
consensus state is $(1-a)b\max (a,b)$. 

\item 3. Finally, the state after the first round may be $(A_{1}B_{1}G_{2})$%
, namely both Bob and Greg change opinion while being exposed to Alice's
perspective. This happens with probability $ab$. The targeted consensus
state remains $(A_{1}B_{1}G_{1})$, and the facilitator has for only option
to expose Greg to Bob's perspective, offering him the opportunity to turn
his opinion back to $G_{1}$. This happens with probability $a^{\prime }$, so
that the overall probability for this scenario is $aba^{\prime }$.
\end{enumerate}

In summary, three of the above scenarios end up in the consensus state $%
(A_{1}B_{1}G_{1})$, with total probability $p_{1}=a(1-b)+(1-a)(1-b)a^{\prime
}+aba^{\prime }$. While the probability to reach consensus state $%
(A_{2}B_{2}G_{2})$ is given by $p_{2}=(1-a)b\max (a,b)$. The total
probability to reach consensus over two rounds when both Bob and Greg are
active in the first round is hence $p_{1}+p_{2}$. Recall that if instead
only Bob is active in the first round, the probability to reach consensus
after two rounds is $p_{0}=a+(1-a)a^{\prime }$.\\ 

\textit{Comparison of both scenarios.--}
We now investigate the possibility that $p_{1}+p_{2}\geq p_{0}$. To simplify the analysis, we consider the case $a=a^{\prime }$, namely Bob's perspective is as much correlated with Alice's than with Greg's. We first consider
and $b>a$, namely the Alice-Bob and Bob-Greg correlation (quantified by $a$) is stronger than the Alice-Greg correlation (quantified by $b$). A simple calculation shows that in this case, $p_{1}+p_{2}=p_{0}+b(1-a)(b-2a)$. As both $b\geq 0$ and $1-a\geq 0$, we
conclude that $p_{1}+p_{2}\geq p_{0}$ if and only if $b\geq 2a$. When $b<a$ we always find that $p_0 \ge p_1 + p_2$. In summary, when $a=a'$, it is optimal to insist twice on making Bob change opinion if $0 \le b \le 2a$; while if $b \ge 2a$, it is optimal to have both Bob and Greg active in the first round, allowing to reach any of the consensus states $(A_1 B_1 G_1)$ or $(A_2 B_2 G_2)$. 

\end{document}